\documentclass[letterpaper, 10pt, conference]{ieeeconf}
\usepackage[top=0.75in, bottom=1.03in, left=0.625in, right=0.759in]{geometry}
\IEEEoverridecommandlockouts

\usepackage{cite}
\usepackage{amsmath,amssymb,amsfonts}
\usepackage{algorithmic}
\usepackage{graphicx}
\usepackage{textcomp}
\usepackage{xcolor}
\usepackage{algorithm}
\usepackage{stfloats}
\usepackage{booktabs}
\usepackage{multirow}
\usepackage{color}
\usepackage[caption=false,font=normalsize,labelfont=sf,textfont=sf]{subfig}
\usepackage{marvosym} 
\usepackage[T1]{fontenc}

\title{\LARGE \bf
SEIDM: A Safe and Efficient Intelligent Driver Model for Autonomous Driving Behavior
}

\begin{document}


\author{
    Yuyang Yao$^{1}$\thanks{$^{1}$Yuyang Yao is with the Department of Information and Communication Engineering, Tongji University, Shanghai, China, \texttt{ 2331814@tongji.edu.cn}.} and
    Shaocheng Luo$^{2}$\thanks{$^{2}$Shaocheng Luo is with the Department of Electrical and Computer Engineering, Duke University, Durham, NC, USA.}
    \thanks{Corresponding Author: S. Luo, \texttt{shaocheng.luo@duke.edu}.}
}

\maketitle
\thispagestyle{empty}
\pagestyle{empty}

\maketitle

\begin{abstract}
The Intelligent Driver Model (IDM) is a cornerstone of Adaptive Cruise Control (ACC), valued for its interpretable parameters and effectiveness in car-following behavior modeling. However, its inherent conservatism leads to prolonged stabilization and reduced traffic efficiency, which have received limited attention. In this paper, we propose SEIDM (Safe and Efficient Intelligent Driver Model), an enhanced IDM extension designed to improve traffic flow efficiency without sacrificing safety. SEIDM introduces an adaptive safety factor to dynamically modulate the impact of the safe deceleration term in acceleration decisions. This allows vehicles to follow more assertively under safe conditions while behaving more cautiously in potential hazards. Extensive urban traffic simulations show that SEIDM achieves significantly shorter stabilization spacing and faster convergence to traffic flow equilibrium, outperforming the original IDM and its variants in traffic stability and efficiency.
\end{abstract}


\section{Introduction}
In the automotive industry, Advanced Driver Assistance Systems (ADAS) have been widely adopted to enhance driving safety and comfort. As a core component of ADAS, Adaptive Cruise Control (ACC) enables vehicles to autonomously adjust their speed in response to traffic conditions~\cite{ref1}. At the heart of ACC systems lies the Intelligent Driver Model (IDM), a widely adopted and influential framework for modeling longitudinal vehicle behavior.

Since its introduction by Treiber et al.~\cite{ref3}, IDM has become pivotal in both academic research and commercial ACC applications. Its popularity stems from an optimal balance of interpretability, simplicity, and effectiveness. With parameters that have clear physical meanings, IDM facilitates straightforward tuning and robust deployment across diverse traffic scenarios. 

Over the past decades, ACC technologies have undergone significant evolution, beginning with the first ACC-equipped vehicle in 1995 and remaining a key focus for automotive leaders, like Audi. Meanwhile, IDM has been extensively refined to address practical challenges, including handling abnormal driving situations, improving parameter sensitivity, and enabling coordination in connected vehicle environments. Given its foundational role, advancing IDM is critical for the development of the next-generation autonomous driving.

To address anomalous vehicle velocities, such as negative values or divergence to $-\infty$ in finite time, Dardour et al.\cite{ref4} incorporated discontinuous dynamics. Ming et al.\cite{ref5} improved parameter calibration by considering heterogeneous driver reaction times and multi-class vehicle dynamics. Liu et al.\cite{ref8}, Wang et al.\cite{ref7}, and Zhou et al.\cite{ref12} incorporated the influence of non-adjacent leading vehicles.

While existing studies have contributed valuable improvements to IDM, most overlook a critical flaw: inherent inefficiency. In practice, IDM often exhibits overly conservative acceleration behaviors. This results in excessive equilibrium spacing, frequent deceleration, and slow traffic flow stabilization–a significant challenge in urban environments where throughput maximization is paramount.

To overcome this, we propose the Safe and Efficient Intelligent Driver Model (SEIDM), which maintains rigorous safety guarantees while significantly improving model efficiency. SEIDM introduces a dynamic risk factor that evaluates car-following safety by jointly considering Time-to-Collision (TTC) and Time Headway (TH) relative to safety thresholds. This factor adaptively modulates the contribution of the safe interaction deceleration term in the IDM's formula.  This allows for more assertive acceleration when safe and stronger deceleration when dangerous. As a result, SEIDM allows the followers to reach the desired velocity in a shorter period, thereby preventing the inter-vehicle spacing from further expansion. Additionally, SEIDM reduces overall traffic flow stabilization period and improves the traffic throughput.

The main contributions of this paper are as follows:
\begin{itemize}
\item We propose SEIDM, a novel extension of the Intelligent Driver Model that improves traffic flow efficiency and speeds up convergence to equilibrium, while maintaining strict safety. This promotes safer and more efficient transportation, benefiting urban mobility and autonomous driving systems.
\item We introduce a dynamic risk assessment mechanism that adaptively balances Time-to-Collision (TTC) and Time Headway (TH) against safety thresholds, enabling real-time, risk-aware acceleration control.
\end{itemize}

The rest of this paper is organized as follows. Section~\ref{sec:related work} reviews related work on IDM and its extensions. Section~\ref{sec:problem} discusses the limitations of the original IDM. The proposed SEIDM model is detailed in Section~\ref{sec:methods}. Section~\ref{sec:experiment setting} outlines the experimental setup. Results and analysis are presented in Section~\ref{sec:results}. Finally, Section~\ref{sec:conclusion} concludes the paper.

\section{Related Work}
\label{sec:related work}
{\bf IDM and Its Variants:} After proposed in 2000, IDM has been extensively refined. To correct unrealistic vehicle behavior, some revisions incorporate discontinuity traits to prevent reverse motion and maintain a minimum following gap~\cite{ref4,ref26}. Derbel et al.~\cite{ref1,ref2} addressed excessive deceleration in emergency scenarios by adding a new term $cv^2/b$ to the expected safe distance formula. Treiber and Helbing~\cite{ref15} incorporate memory effects in IDM. Treiber and Kesting~\cite{ref16} add external noise and action points to study mechanisms behind traffic flow instabilities and indifferent regions of finite human perception thresholds. Xiao et al.~\cite{ref17} propose IDM-GARCH model, which is based on the uncertainty of velocity fluctuation caused by drivers' perceptual uncertainty. To build a realistic agent model, Eggert et al.~\cite{ref22} propose the Foresighted Driver Model, balancing the predictive risk with utility.

{\bf IDM Parameter Optimization:} A key area of development involves adapting IDM parameters for heterogeneity and dynamic conditions. Studies have explicitly integrated driver reaction times and vehicle-specific parameters to model different vehicle types more accurately~\cite{ref4,ref5}. To move beyond static parameters, self-adaptive mechanisms dynamically adjust acceleration based on real-time inputs like the leading vehicle's speed and TH~\cite{ref9}. Further frameworks enable context-dependent acceleration bounds~\cite{ref10} and desired speed adjustments based on individual driving habits and traffic conditions~\cite{ref11, ref25}.

\section{Problem Statement}
\label{sec:problem}
\subsection{Classic IDM}
\begin{figure}[!ht]
  \centering
  \includegraphics[width=0.47\textwidth]{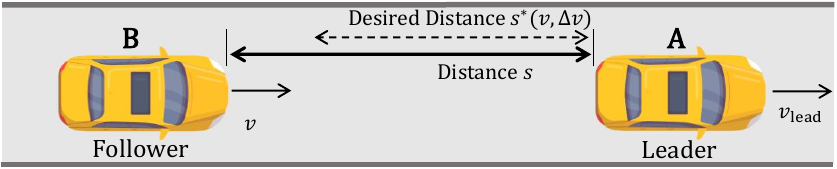}
  \caption{Representation of IDM. Vehicle B continuously adjusts its acceleration based on its and Vehicle A's motion.}
  \label{fig1}
\end{figure}

IDM controls the following behavior based on the distance and velocities of the leader and follower, shown in Figure~\ref{fig1}.

The calculation of acceleration can be described as:
\begin{equation}
    \label{acceleration calculation}
    a_{\rm IDM}(s, v, \Delta v)=a_{0}\left[1-\left(\frac{v}{v_{0}}\right)^{\delta}-\left(\frac{s^{*}(v, \Delta v)}{s}\right)^{2}\right], 
\end{equation}
where $v$ is the follower's current speed, $s$ is the headway distance, and $\Delta v$ is the speed difference. The model parameters are: maximum acceleration $a_0$, desired free-flow speed $v_0$, and acceleration exponent $\delta$, which indicates how quickly acceleration slows down in relation to the speed. The desired safety distance $s^{*}(v, \Delta v)$ is calculated as:

\begin{equation}
    \label{acceleration calculation}
    s^{*}(v, \Delta v)=s_{0}+\max \left(0, v T+\frac{v \Delta v}{2 \sqrt{a_{0} b_{0}}}\right), 
\end{equation}
where $s_0$ is the minimum gap under static conditions, $T$ is the safe TH, and $b_0$ represents the comfortable deceleration. 

The term $\mathcal{A} = 1-\left(\frac{v}{v_{0}}\right)^{\delta}$ governs free-flow acceleration, which decreases as the follower's velocity $v$ approaches the desired speed $v_0$. The safe interaction deceleration term $\mathcal{D} = (\frac{s^{*}(v, \Delta v)}{s})^{2}$ regulates safe following. When $s > s^{*}(v, \Delta v)$, the term diminishes and encourages the follower to accelerate to reduce the gap. Otherwise, the term increases and forces deceleration to restore a safe distance.

\subsection{Conservativeness of IDM}
\begin{figure}[!ht]
  \centering
  \includegraphics[width=0.47\textwidth]{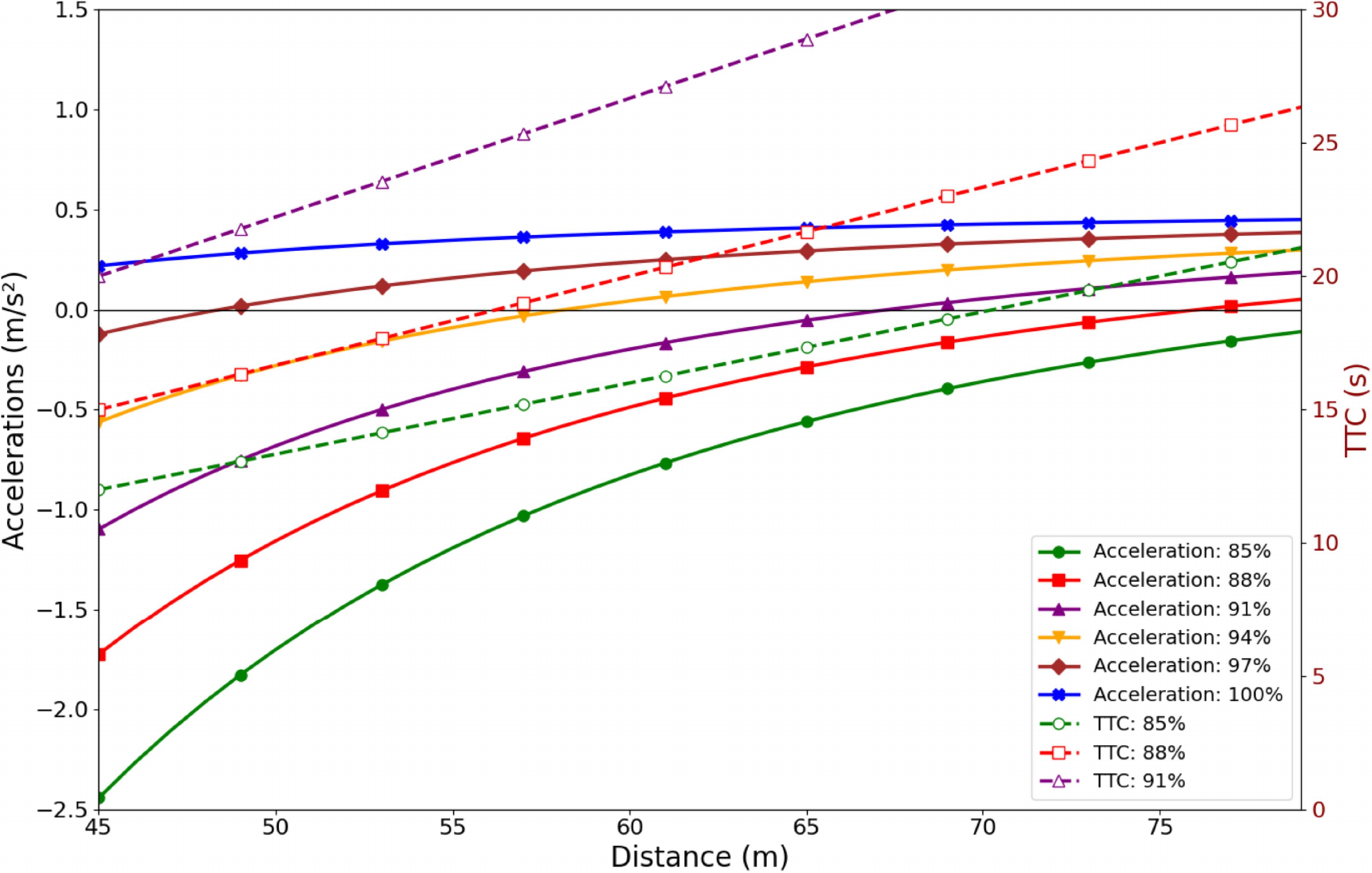}
  \caption{The illustration of follower acceleration versus inter-vehicle distance across a range of relative speeds, with corresponding TTC values shown on the right axis. In this analysis, the follower's speed is constant at 90 $\mathrm {km/h}$, while the leader's speed varies from $85\%$ to $100\%$ of this value. The IDM parameters from Section~\ref{Model Parameter Configuration} are used. }
  \label{fig8}
\end{figure}

\begin{figure}[!ht]
  \centering
  \includegraphics[width=0.47\textwidth]{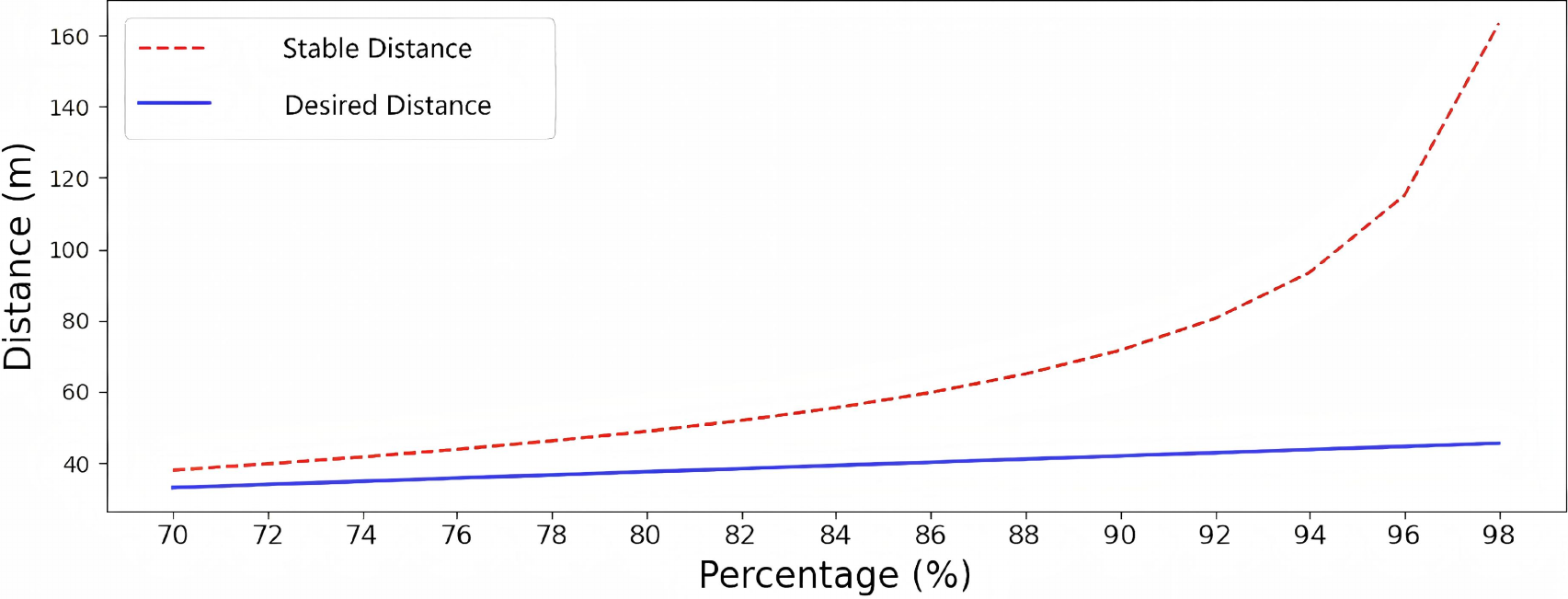}
  \caption{Comparison between the stabilized following distance and the desired distance in IDM. The horizontal axis represents the ratio of the stabilized speed to the desired speed $v_0$. As the stabilized speed approaches $v_0$, the gap between the actual and desired distances increases markedly, highlighting IDM's growing deviation from optimal spacing at higher speeds.}
  \label{fig7}
\end{figure}

In car-following scenarios, IDM exhibits overly conservative behavior, generating low or negative acceleration even under safe conditions with high Time-to-Collision (TTC > 10), as shown in Figure~\ref{fig8}. This unnecessary caution, with accelerations typically below $0.25~ \mathrm{m/s}^2$, delays the follower's acceleration until the leader reaches its desired speed, creating unnecessarily large inter-vehicle gaps. In particular, the gap between the stabilized distance and the desired distance $s^{*}(v, \Delta v)$ increases as the follower reaches the desired speed $v_0$, shown in Figure~\ref{fig7}.

This conservative behavior scales to diminish overall traffic throughput, exacerbating congestion in dense urban networks. The prolonged stabilization period reflects a delayed convergence to steady-state flow and inefficient road capacity usage. To address this, we propose SEIDM, which enables more responsive acceleration under safe conditions. This allows followers to attain their desired velocity faster, shortening inter-vehicle gaps, accelerating traffic stabilization, and ultimately increasing overall traffic throughput and efficiency.

\section{Methods}
\label{sec:methods}
\subsection{Risk Factor}
We define a $risk\_factor$ that integrates two critical safety metrics for car-following risk assessment:
\begin{itemize}
\item $TTC$: Time-to-Collision, which measures temporal proximity to potential collision.
\item $TH$: Time Headway, which represents the temporal spacing between vehicles. 
\end{itemize}

The $risk\_factor$ is formulated by jointly evaluating these metrics relative to respective safety thresholds:
\begin{equation}
\label{risk parameter}
    risk\_factor=\left\{\begin{array}{c}
\frac{T}{TH} \quad \text { if } \frac{TTC_{0}}{TTC}<\frac{T}{TH}-\varepsilon \\[0.6em]
f(\frac{TTC_{0}}{TTC},\frac{T}{TH}) \text { if }\left|\frac{TTC_{0}}{TTC}-\frac{T}{TH}\right| \leq \varepsilon \\[0.6em]
\frac{TTC_{0}}{TTC} \quad \text { if } \frac{TTC_{0}}{TTC}>\frac{T}{TH}+\varepsilon
\end{array}\right.,
\end{equation}
    
\begin{equation}
\label{transition}
f(\frac{TTC_{0}}{TTC},\frac{T}{TH}) = \alpha \cdot \frac{TTC_{0}}{TTC}+(1-\alpha) \cdot \frac{T}{TH},
\end{equation}

\begin{equation}
\label{alpha}
\alpha=\frac{1}{2}+\frac{1}{2 \varepsilon}\left(\frac{TTC_{0}}{TTC}-\frac{T}{TH}\right),
\end{equation}
where $TTC_0 = 2.7s$ is the safe collision threshold from the Commercial Vehicle Driving Standard (JT/T 883)~\cite{ref18}, and $T$ is IDM's safe TH. To smooth the two-stage transition, we introduce an intermediate phase $f(\frac{TTC_{0}}{TTC},\frac{T}{TH})$, which is a weighted sum of $\frac{TTC_{0}}{TTC}$ and $\frac{T}{TH}$ defined in Equation~\ref{transition}. $\alpha \in [0,1]$ denotes the weighting coefficient. The transition duration $\varepsilon = 0.1\cdot\frac{T}{TH}$ balances mutation suppression with segment logic validity~\cite{ref24}.

From the above equations, $risk\_factor$ evaluates two types of car-following risk: imminent collision potential (based on $TTC$) and insufficient following distance that may necessitate emergency braking (based on $TH$). When $risk\_factor > 1$, it means dangerous behavior, requiring the follower to decelerate and increase the gap; otherwise, the state is safe, allowing potential acceleration.

The rationale for taking the maximum of $\frac{TTC_0}{TTC}$ and $\frac{T}{TH}$ stems from the inherent complexity and variability of traffic situations, where no single criterion can capture all hazardous scenarios. While $\frac{TTC}{TTC_0}$ serves as the main indicator, this metric alone fails to adequately assess danger in high-speed car-following scenarios with minimal speed differences between consecutive vehicles. In this situation, $TTC$ remains large( $\frac{TTC_{0}}{TTC} << 1$) even as the inter-vehicle gap decreases, thereby wrongly indicating safe conditions. However, in such situations, $\frac{T}{TH}$ becomes the effective indicator ($\frac{T}{TH} > 1 >> \frac{TTC_{0}}{TTC}$). Thus, we employ $\frac{T}{TH}$ as a complementary hazard condition indicator. 

\subsection{SEIDM}
We utilize $risk\_factor$ to adjust the influence of the safe interaction deceleration term on the acceleration: 
\begin{equation}
\label{SEIDM}
\begin{split}
a_{\rm SEIDM}(s, v, \Delta v) = & \,a_{0}\cdot\mathcal{A} - a_{0}\cdot{risk\_factor}^{r} \cdot\mathcal{D},
\end{split}
\end{equation}
where $\mathcal{A}$ and $\mathcal{D}$ represent the free-flow acceleration and safe interaction deceleration terms, respectively. The risk response indicator $r$ calibrates the model's sensitivity to the ${risk\_factor}$, allowing for adaptive behavior across different scenarios. A higher value of $r$ intensifies the influence of perceived risk on the deceleration term, resulting in stricter braking responses during critical events.

\begin{figure}[!ht]
  \centering
  \includegraphics[width=0.47\textwidth]{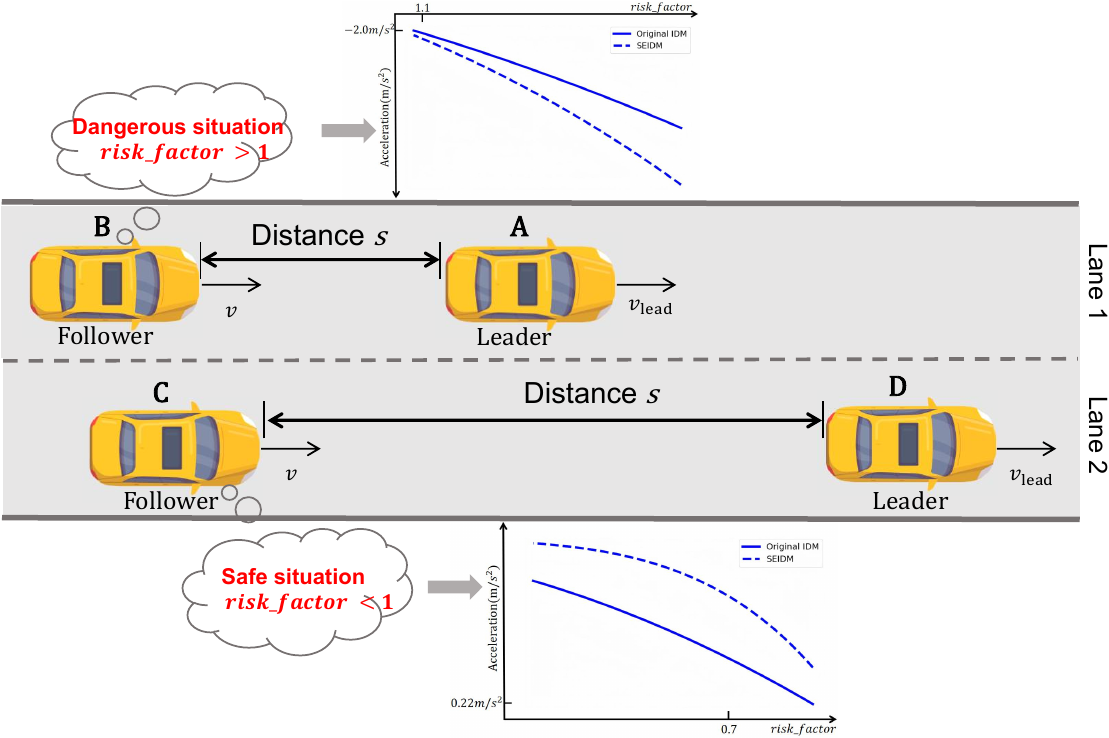}
  \caption{Illustration of SEIDM behavior. Compared to IDM (solid lines), SEIDM (dashed lines) provides stronger deceleration for quicker hazard response (Vehicle B → Vehicle A) and more aggressive acceleration to reach the desired speed efficiently under safe conditions (Vehicle C → Vehicle D). }
\label{fig6}
\end{figure}

From the above equation, the safe interaction deceleration is proportional to $risk\_factor$. Unsafe behavior triggers greater speed reduction than IDM to rapidly extend following distance, while safe conditions allow acceleration to the desired or maximum speed. This is illustrated in Figure~\ref{fig6}.

\subsection{Stability Analysis}
This section analyzes the stability of SEIDM, showing that the model inherently prevents unsafe speeds. Speed is primarily limited by $\mathcal{A}$: as the follower's speed $v$ approaches the desired speed $v_0$, the term $\left(\frac{v}{v_{0}}\right)^{\delta}$ approaches 1, causing $\mathcal{A}$ to approach zero. This ensures asymptotic convergence toward $v_0$. Furthermore, if $v$ exceeds $v_0$, $\mathcal{A}$ becomes negative, forcing immediate deceleration back to the desired speed. This stability is reinforced by ${risk\_factor}^{r} \cdot \mathcal{D}$, which adds an extra deceleration effect. Consequently, SEIDM bounds all vehicle speeds by a certain upper limit, guaranteeing a rational and safe speed range.

\subsection{Computational Efficiency}
It is worth noting that SEIDM retains the computational simplicity of the original IDM. The calculation of the $risk_factor$ involves only basic arithmetic operations and conditional checks, maintaining a time complexity of $O(1)$.

\section{Experiment setting}
\label{sec:experiment setting}
\subsection{Experimental Scenarios}
\label{Experimental Scenarios}
There are four distinct traffic scenarios in the experiment:

{\bf Scenario~\uppercase\expandafter{\romannumeral1}: Initial chaotic traffic flow stabilization scenario.} In a double-lane highway, all vehicles start with random speeds ($80\%-100\%$ of the highway's limited speed $v_{\rm max}$) and the initial spacing matches the desired distance prescribed by IDM. The scenario observes the system's convergence to equilibrium spacing and $v_{\rm max}$.  

\begin{figure}[!ht]
  \centering
  \includegraphics[width=0.40\textwidth]{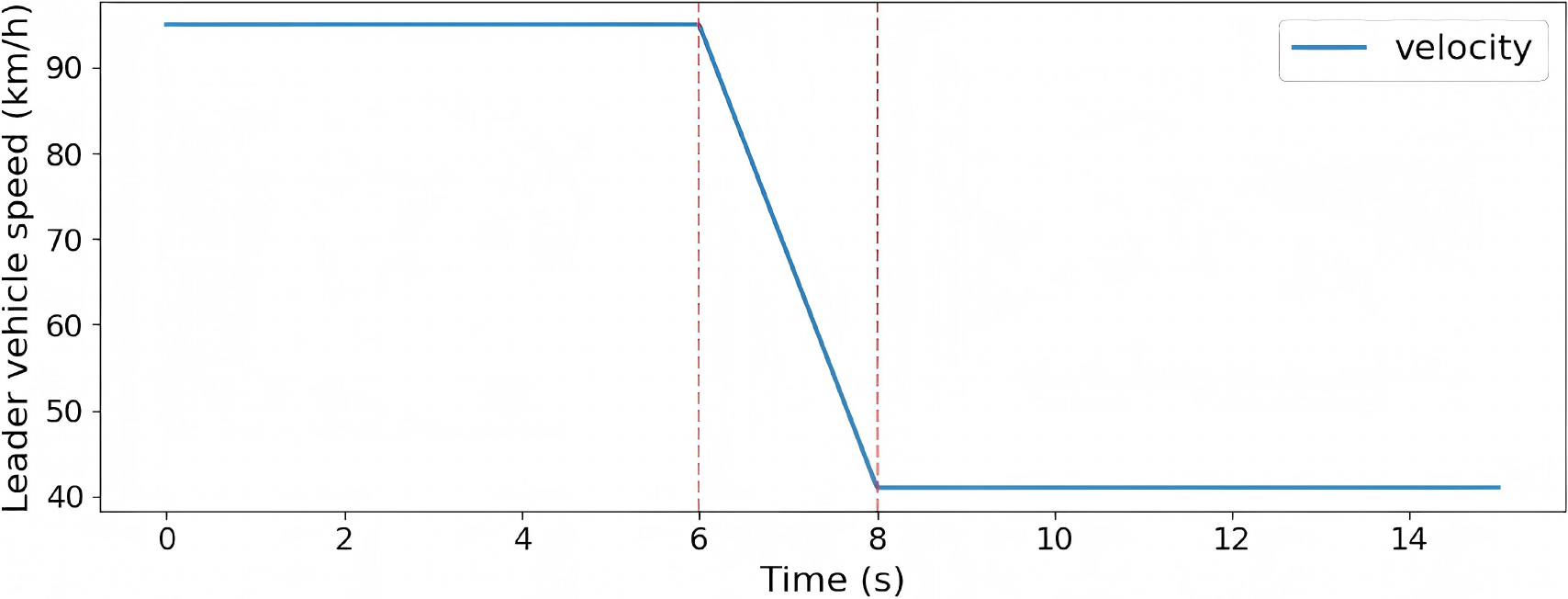}
  \caption{Leader speed profile.}
  \label{fig2}
\end{figure}

{\bf Scenario~\uppercase\expandafter{\romannumeral2}: Consecutive vehicle emergency braking scenario.} Two vehicles travel at $v_{\bf max}$, maintaining an equilibrium spacing. The leader decelerates suddenly. The leader executes a sudden deceleration (Figure~\ref{fig2}), testing the follower's collision-avoidance response~\cite{ref2}. 

{\bf Scenario~\uppercase\expandafter{\romannumeral3}: 
Traffic Flow Emergency Braking Scenario.} 
A single-lane traffic flow travels at $v_{\bf max}$ with equilibrium spacing. The lead vehicle decelerates suddenly (Figure~\ref{fig2}), testing collision-avoidance propagation in the traffic flow.

{\bf Scenario~\uppercase\expandafter{\romannumeral4}: Equilibrium traffic flow with vehicle insertion scenario.} In a stable double-lane flow at $v_{\rm max}$, new vehicles are randomly inserted into two consecutive vehicles per lane with $v_{\rm max}$. The scenario measures how the system absorbs this disturbance and restores equilibrium.

\subsection{Model Parameter Configuration}
\label{Model Parameter Configuration}
The parameter configurations for IDM and SEIDM are presented in Table~\ref{params}. In order to model realistic behavior, reaction time $T'$ is incorporated into the car-following model~\cite{ref8}.

\begin{table}[t]
\centering
\caption{IDM's and SEIDM's Parameter Configuration}
\vspace{-2.5mm}
\label{params}
\renewcommand{\arraystretch}{1.1} 
\setlength{\tabcolsep}{8pt} 
\begin{tabular}{c c c c}
\hline
    Parameter & Variable & Value & Unit \\
    \hline
    Maximum acceleration & $a_0$ & 1.46 & $\mathrm {m/s^2}$ \\
    Comfortable deceleration & $b_0$ & 2.0 & $ \mathrm {m/s^2}$ \\
    Desired speed & $v_{0}$ & 100 & $\mathrm {km/h} $ \\
    Limited maximum speed & $v_{\rm max}$ & 95 & $\mathrm {km/h}$ \\
    Safe time headway & $T$ & 1.6 & $\mathrm s$ \\
    Static required distance & $s_0$ & 2 & $\mathrm m$ \\
    Acceleration exponent & $\delta$ & 4 & -- \\
    Reaction time & $T'$ & 1.0 & $\mathrm s$ \\
    Safe collision threshold & $TTC_0$ & 2.7 & $\mathrm s$ \\
    \hline
\end{tabular}
\end{table}

\section{Results and Discussions}
\label{sec:results}
This section presents experimental evaluations conducted in the CARLA simulator. In Section~\ref{Selection}, we perform the comparative experiment to determine the risk response indicator $r$. In Section~\ref{scene 1}, \ref{scene 2}, and \ref{scene 3}, we compare SEIDM with other common car-following models in four critical scenarios described in Section~\ref{Experimental Scenarios}. 

\subsection{Selection of Risk Response Indicator}
\label{Selection}
The parameter $r$ is evaluated in two key scenarios: traffic flow stabilization and emergency braking. In the stabilization scenario, the final stabilization spacing and the resulting traffic throughput are recorded, as these metrics reflect the model's efficiency~\cite{ref23}. A shorter stabilization spacing indicates faster convergence and higher throughput. In the emergency braking scenario, the braking duration and peak deceleration are measured to assess the smoothness and safety of the braking response~\cite{ref4}. The experimental results for varying values of $r$ are summarized in Table~\ref{different paramter}.

\begin{table}[!ht]
\centering
\caption{Results on the Selection of $r$}
\vspace{-2.5mm}
    \label{different paramter}
    \renewcommand{\arraystretch}{1.1}
    \setlength{\tabcolsep}{6pt}
		\begin{tabular}{c c c c c}
            \hline
            $r$ & $0$ & $0.05$ & $0.1$ & $0.2$\\
            \hline
            Stabilization Spacing ($\mathrm m$) & $102.67$ & $100.47$ & $98.41$ & $94.70$ \\
            Traffic Throughput ($\mathrm {veh/h}$) & $925.3$ & $945.5$ & $965.3$ & $1003.1$ \\
            Braking Duration ($\mathrm s$) & $31.6$ & $29.3$ & $26.95$ & $25.8$ \\
            Peak Deceleration ($\mathrm {m/s^2}$) & $-4.05$ & $-4.14$ & $-4.2$ & $-4.31$ \\
            \hline
            $r$ & $0.3$ & $0.4$ & $0.5$ & $0.6$\\
            \hline
            Stabilization Spacing ($\mathrm m$) & $91.43$ & $88.53$ & $85.95$ & $83.64$ \\
            Traffic Throughput ($\mathrm {veh/h}$) & $1039.0$ & $1073.1$ & $1105.3$ & $1135.8$ \\
            Braking Duration ($\mathrm s$) & $24.5$ & $23.5$ & $22.4$ & $21.15$ \\
            Peak Deceleration ($\mathrm {m/s^2}$) & $-4.42$ & $-4.54$ & $-4.66$ & $-4.79$ \\
            \hline
            $r$ & $0.7$ & $0.8$ & $0.9$ & $1.0$\\
            \hline
            Stabilization Spacing ($\mathrm m$) & $81.54$ & $79.64$ & $77.92$ & $76.34$ \\
            Traffic Throughput ($\mathrm {veh/h}$) & $1165.1$ & $1192.9$ & $1219.2$ & $1244.4$ \\
            Braking Duration ($\mathrm s$) & $20.2$ & $19.2$ & $18.1$ & $16.7$ \\
            Peak Deceleration ($\mathrm {m/s^2}$) & $-4.92$ & $-5.06$ & $-5.19$ & $-5.32$ \\
            \hline
        \end{tabular}
\end{table}

As $r$ increases from 0 to 1, the stabilization spacing decreases from 102.67 $\mathrm m$ to 76.34 $\mathrm m$, while traffic throughput improves by 34.6\% (from 925.3 $\mathrm {veh/h}$ to 1244.4 $\mathrm {veh/h}$). In contrast, the braking duration shortens by 47.2\%, and peak deceleration becomes more intense, rising from –4.05 $\mathrm {m/s^2}$ to –5.32 $\mathrm {m/s^2}$. These results reveal a clear trade-off: higher $r$ enhance traffic efficiency and throughput at the expense of more abrupt braking.

Based on these observations, SEIDM operation can be classified into three distinct regimes:
\begin{itemize}
\item {\bf Smoothness-priority regime} ($r \in (0, 0.4]$): prioritizes ride comfort over traffic efficiency.
\item {\bf Balanced regime} ($r \in [0.4, 0.8]$): achieves a trade-off between comfort and efficiency.
\item {\bf Efficiency-priority regime} ($r \in (0.8, \infty]$): maximizes throughput, though with increased braking intensity.
\end{itemize}

We utilize the balanced parameter ($r = 0.6$) for subsequent comparative trials. Verification against the ISO 15622 standard confirms that the system's average peak deceleration ($2~\mathrm s$ period) remains well within the acceptable safety range. 

\subsection{Comparison Experiments in Scenario~\uppercase\expandafter{\romannumeral1}}
\label{scene 1}
We compare SEIDM with several established car-following models, including the Krauss model (Krauss-FM)~\cite{ref19}, the original IDM, and its variants—in Scenario~\uppercase\expandafter{\romannumeral1}. The simulation uses 40 vehicles per lane, and we evaluate each model over 20 independent trials. For each trial, we record the stabilization spacing, stabilization time, and traffic throughput, with average values reported. The comparison results are summarized in Table~\ref{comparison in stabilization scenario}.

\begin{table}[!ht]
\centering
\caption{Experimental Results in Traffic Flow Stabilization Scenario}
\vspace{-2.5mm}
\label{comparison in stabilization scenario}
\renewcommand{\arraystretch}{1.1}
    \setlength{\tabcolsep}{3pt}
    \begin{tabular}{c c c c}
        \hline
        Model & Spacing ($\mathrm m$) & Period ($\mathrm s$) & Traffic Throughput ($\mathrm {veh/h}$) \\
        \hline
        Krauss-FM~\cite{ref19} & $44.28$ & $3.4$ & $2145.44$ \\
        IDM~\cite{ref3} & $102.67$ & $2472.4$ & $925.29$ \\
        Deng et al.~\cite{ref20} & $102.67$ & $2483.2$ & $925.29$ \\
        Derbel et al.~\cite{ref1} & - & - & - \\
        Dardour et al.~\cite{ref4} & $196.06$ & $4490.6$ & $484.55$ \\
        {\bf SEIDM} & $83.64$ & $1479.8$ & $1135.8$ \\
        \hline
        \end{tabular}
\end{table}

\begin{figure}[!ht]
  \centering
  \includegraphics[width=0.30\textwidth]{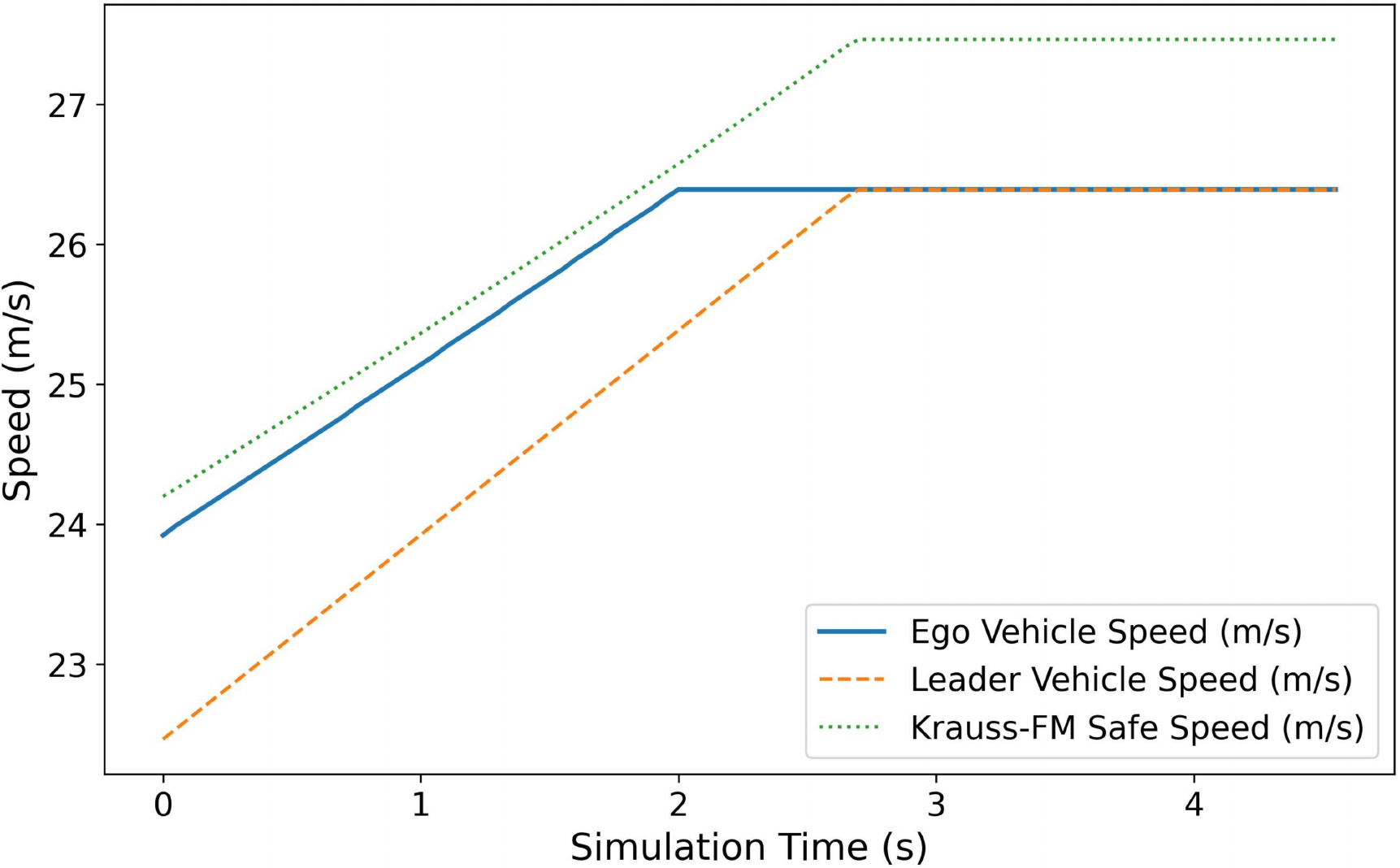}
  \caption{Relationship among the ego velocity, the leader velocity, and the safe velocity of a vehicle in traffic flow.}
  \label{fig3}
\vspace{-5mm}
\end{figure}

Among the evaluated car-following models, two stand out due to anomalous behavior: Krauss-FM and the revised IDM proposed by Derbel et al. Krauss-FM, in particular, shows anomalously high performance, with implausibly short equilibrium times (3.4 $\mathrm s$) and inter-vehicle distances (44.28 $\mathrm m$). This abnormality stems from its defined safe velocity $v_{\rm safe}$, leading to overly aggressive acceleration that violates realistic safety margins.
\begin{equation}
\label{v safe}
\begin{split}
v_{\rm safe}=v_{l}(t)+\frac{s-v_{l}(t) \cdot T'}{\frac{v_{l}(t)+v(t)}{2 b_0}+T'},
\end{split}
\end{equation}
where $v_{l}$ is the velocity of the leader, $v$ is the velocity of the follower, $s$ is the inter-vehicle gap, $T'$ is the reaction time and $b_0$ is the comfortable deceleration. These parameters are consistent with those defined in Section~\ref{Model Parameter Configuration}. The target velocity $v_{\rm target}$ for the next step is defined:
\begin{equation}
\label{v next}
\begin{split}
v_{\rm {target }}=\min \left[v_{\max }, v+a_0t, v_{\rm {safe }}\right],
\end{split}
\end{equation}
where $a_0$ is the maximum acceleration. Figure~\ref{fig3} illustrates the relationship between ego velocity, leader velocity, and computed safe velocity.
Since $v_{\rm safe}$ often exceeds $v + a_0 t$, the follower consistently accelerates at the maximum rate $a_0$ until reaching $v_{\rm max}$. This rigid acceleration strategy results in unrealistic speed convergence. Additionally, in emergency braking scenarios, Krauss-FM demonstrates excessively harsh deceleration, with peaks as high as $-7.01~\mathrm{m/s^2}$, highlighting the model's propensity for severe braking responses.

The modified IDM proposed by Derbel et al. adds a new term $cv^2/b$ to $s^{*}(v, \Delta v)$, where $c = 0.4$. In the stabilization scenario, $cv^2/b$ is extremely high, imposing a severe safe interaction deceleration value. As a result, the model exhibits abnormal behavior: only leading vehicles maintain movement while the majority remain stationary. 

\begin{figure}[!ht]
  \centering
  \includegraphics[width=0.48\textwidth]{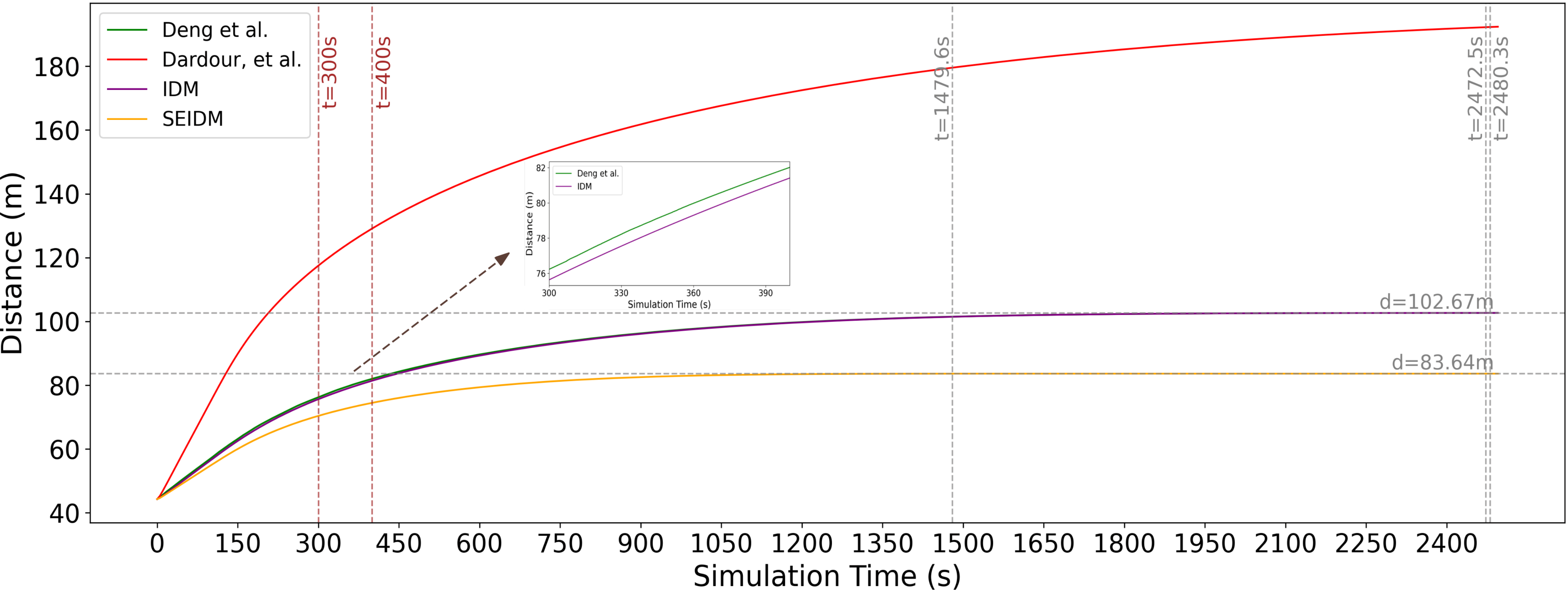}
  \caption{Relationship between average distance and simulation time across various car-following models.}
  \label{fig5}
  \vspace{-5mm}
\end{figure}

SEIDM outperforms existing car-following models, achieving the shortest stabilization spacing (83.64 $\mathrm m$), fastest stabilization time (1479.8 $\mathrm s$), and highest traffic throughput (1135.8 $\mathrm {veh/h}$). As shown in Figure~\ref{fig5}, SEIDM reaches flow stabilization at 83.64 $\mathrm m$ within 1479.6 $\mathrm s$, while the original IDM and Deng et al.'s revised IDM require significantly longer times—2472.5 $\mathrm s$ and 2480.3 $\mathrm s$, respectively—under identical initial conditions for fair comparison.

\subsection{Comparison Experiments in Scenario~\uppercase\expandafter{\romannumeral2} and Scenario~\uppercase\expandafter{\romannumeral3}}
\label{scene 2}
To evaluate the safety performance of SEIDM, we conduct a comparative analysis against the original IDM under emergency braking scenarios. In the two-vehicle case, both vehicles travel at the $v_{\rm max}$, maintaining their respective stabilization distances (IDM: 102.67 $\mathrm m$; SEIDM: 83.64 $\mathrm m$). We record the final inter-vehicle spacing and the total spacing reduction. As shown in Table~\ref{comparison in braking scenario}, SEIDM achieves a final spacing of 17.45 $\mathrm m$, slightly shorter than IDM's, but still within the safety margin defined by ISO 15622. By dynamically adjusting to braking urgency via the $risk\_factor$, SEIDM enables the following vehicle to decelerate more timely, resulting in a smaller distance loss (66.19 $\mathrm m$). These results demonstrate that SEIDM offers enhanced responsiveness in critical situations, compared to IDM. 

\begin{table}[!ht]
\centering
\caption{Experimental Results in the Consecutive Vehicle Emergency Braking Scenario}
\vspace{-3mm}
        \label{comparison in braking scenario}
        \renewcommand{\arraystretch}{1.1}
        \setlength{\tabcolsep}{7pt}
		\begin{tabular}{c c c}
        \hline
        Model & Final Spacing ($\mathrm m$) & Spacing Reduction ($\mathrm m$) \\
        \hline
        IDM~\cite{ref3} & $17.96$ & $84.71$ \\
        {\bf SEIDM} & $17.45$ & $66.19$ \\
        \hline
      \end{tabular}
      \vspace{-3mm}
\end{table}

In a full traffic flow simulation with 10 vehicles at $v_{\rm max}$, SEIDM ensures prompt deceleration and safe gap maintenance when the leading vehicle (Vehicle 9) brakes suddenly, as detailed in Table~\ref{comparison in braking scenario multi}.

\begin{table}[!ht]
    \centering
\caption{Experimental Results in the Traffic Flow Emergency Braking Scenario}
\vspace{-2.5mm}
    \label{comparison in braking scenario multi}
    \renewcommand{\arraystretch}{1.1}
    \setlength{\tabcolsep}{4pt}
		\begin{tabular}{c c c c c c c}
        \toprule
        {Model} & {Metric} & 0 & 1 & 2 & 3 & 4\\
        \midrule
        \multirow{2}{*}{IDM~\cite{ref3}} 
        & Final Spacing ($\mathrm m$) & 17.98 & 17.98 & 17.98 & 17.97 & 17.97 \\
        & Spacing Reduction ($\mathrm m$) & 84.69 & 84.69 & 84.69 & 84.70 & 84.70 \\
        \midrule
        \multirow{2}{*}{\bf SEIDM}
        & Final Spacing ($\mathrm m$) & 17.47 & 17.47 & 17.47 & 17.46 & 17.46 \\
        & Spacing Reduction ($\mathrm m$) & 66.17 & 66.17 & 66.17 & 66.18 & 66.18 \\
        \toprule
        Model & Metric & 5 & 6 & 7 & 8 & 9\\
        \midrule
        \multirow{2}{*}{IDM~\cite{ref3}} 
        & Final Spacing ($\mathrm m$) & 17.70 & 17.970 & 17.97 & 17.96& - \\
        & Spacing Reduction ($\mathrm m$) & 84.70 & 84.70 & 84.70 & 84.71& - \\
        \midrule
        \multirow{2}{*}{\bf SEIDM}
        & Final Spacing ($\mathrm m$) & 17.46 & 17.45 & 17.45 & 17.45& - \\
        & Spacing Reduction ($\mathrm m$) & 66.18 & 66.19 & 66.19 & 66.19& - \\
        \bottomrule
      \end{tabular}
      \vspace{-5mm}
\end{table}

\subsection{Comparison Experiments in Scenario~\uppercase\expandafter{\romannumeral4}}
\label{scene 3}
To evaluate SEIDM's disturbance-handling capability, we conduct vehicle insertion experiments with 80 vehicles (40 per lane) in CARLA. With randomized insertion positions, each experiment was repeated 20 times. For each trial, we record the stabilization period and rearmost vehicle response time—the duration for the last vehicle to react to the insertion event. Mean values across all trials are reported in Table~\ref{comparison in insertion scenario}.

\begin{table}[!ht]
\centering
	\caption{Experimental Results in the Vehicle Insertion Scenario}
    \vspace{-2.5mm}
        \label{comparison in insertion scenario}
        \renewcommand{\arraystretch}{1.1}
        \setlength{\tabcolsep}{8pt}
		\begin{tabular}{c c c}
        \hline
        Model & Stabilization Period ($\mathrm s$)& Response Time ($\mathrm s$) \\
        \hline
        IDM~\cite{ref3} & $2024.2$ & $555.9$ \\
        {\bf SEIDM} & $1276.72$ & $317.516$ \\
        \hline
        \end{tabular}
        \vspace{-3mm}
\end{table}

SEIDM enables the traffic flow to accommodate new vehicle insertion more quickly and effectively than IDM, reducing stabilization time to 1276.72 $\mathrm s$ and trailing vehicle response time to 317.516 $\mathrm s$. This demonstrates SEIDM's superior ability to mitigate traffic disturbances and enhance flow stability.

\section{Conclusion}
\label{sec:conclusion}
In this paper, we present an enhanced intelligent driving model (IDM), named SEIDM, which aims to maximize driving efficiency without compromising safety. Building upon the IDM, SEIDM incorporates a dynamic risk factor that evaluates the follower's behavior to establish safety standards. This factor adaptively modulates the safe interaction deceleration, thereby reducing traffic flow stabilization spacing and time while enhancing disturbance-handling capability. Comparative evaluations demonstrate SEIDM's superior performance against existing car-following models. Future work will explore the impact of connected vehicles and develop an adaptive risk-response indicator to improve driving efficiency and traffic throughput, particularly in multi-lane scenarios. 

\bibliographystyle{IEEEtran}
\bibliography{IEEEabrv,reference}

\end{document}